\documentclass[12pt,doublespace]{revtex4}

\usepackage{setspace}

\usepackage{amsfonts}
\usepackage{mathrsfs}
\usepackage{amsmath}
\usepackage{graphicx}
\usepackage{dcolumn}
\usepackage{bm}
\usepackage{amssymb}

\begin{document}

\title{Restoring Coherence Lost to a Slow Interacting Mesoscopic Bath}
\author{Wang Yao}
\thanks{Presently at Department of Physics, The University of Texas, Austin, Texas 78712-0264}
\author{Ren-Bao Liu}
\thanks{Presently at Department of Physics, The Chinese University of Hong Kong, Hong Kong, China}
\author{L. J. Sham}
\affiliation{Department of Physics, University of California San Diego, La Jolla, California 92093-0319}
\date{\today}

\begin{abstract}
For a two-state quantum object interacting with a slow mesoscopic
interacting spin bath, we show that a many-body solution of the bath
dynamics conditioned on the quantum-object state leads to an
efficient control scheme to recover the lost quantum-object
coherence through disentanglement. We demonstrate the theory with
the realistic problem of one electron spin in a bath of many
interacting nuclear spins in a semiconductor quantum dot. The spin
language is easily generalized to a quantum object in contact with a
bath of interacting multi-level quantum units with the caveat that
it is mesoscopic and its dynamics is slow compared with the quantum
object.
\end{abstract}

 \keywords{}
\maketitle

Coherent superposition of states of a quantum object is the
wellspring of quantum properties and key to quantum technology.
Decoherence of a quantum object results from the entanglement with
an environment by coupled dynamics
\cite{Zurek_decoherence_RMP,Decoherence_classicalWorld,Schlosshauer_decoherence}.
Amelioration of decoherence becomes important in any sustained
quantum process. Different types of amelioration include dynamical
decoupling \cite{concatenation_Lidar,DD_VL,Viola_Random},
decoherence-free subspace \cite{DFS}, quantum error correction (for
a review, see \cite{NielsenChuang}), and feedback
control~\cite{Buscemi_invert_decoherence}.

We offer an alternate approach to the restoration of coherence based
on the theory that control of the quantum object can direct the
quantum evolution of the bath to disentangle the object from the
bath. The operation resembles the spin echo schemes
\cite{Slichter_NMR} but it removes the pure decoherence due to bath
interaction dynamics as well as the inhomogeneous broadening effect.
The key is that the environment is effectively a mesoscopic system,
i.e., the number of particles $N$ is small enough for the timescale
of the quantum object decoherence to be much smaller than its energy
relaxation time $T_1$ while large enough for ergodicity,
specifically for the Poincar\'{e} period to be effectively infinite
as compared to $T_1$. The theoretical demonstration of coherence
restoration uses one electron spin in a semiconductor quantum dot of
many ($N\sim 10^6$) nuclear spins, which serves as a paradigmatic
system of a two-level system in a bath of interacting spins for
decoherence physics \cite{Stamp_spinbath} and for spin-based quantum
technology \cite{Loss_QDspinQC,Imamoglu_CQED_Spin}. Electron spin
decoherence due to the hyperfine interaction with the nuclear spins
has been much studied (see \cite{Espin_HF_1_Loss} and references
therein). Theories of the effect of interaction between nuclear
spins on the electron decoherence have recently
appeared~\cite{Espin_HF_3_DasSarma,Espin_YLS}. Our theory of
coherence recovery by disentanglement is based on the previous
finding~\cite{Espin_YLS} that the mesoscopic bath of slow dynamics
is well described by a simple pseudospin model for the particle pair
interaction in the bath.

We start by defining the decoherence process in a simple quantum
prescription. The model for the coupled spin-bath system is  a
localized electron of spin 1/2 coupled to a bath of  finite $N$
mutually interacting nuclei with spin $j$ in a magnetic field. The
isolation of the electron spin plus the meso-bath, in the timescale
of the total processing duration $T_p$, from the rest of universe
arises out of their weak coupling with the outside. The initial
state of the electron spin,
$|\varphi^s(0)\rangle=C_+|+\rangle+C_-|-\rangle$, is prepared as a
coherent superposition of the spin up and down states $|\pm \rangle$
in an external magnetic field. The state of the total system of the
spin plus bath at that instant forms an {\it unentangled} state,
$|\Psi(0)\rangle=|\varphi^s(0)\rangle\otimes|{ \mathcal J}\rangle$.
It evolves over time $t$ to the {\it entangled}
$|\Psi(t)\rangle=C_+(t)|+\rangle\otimes|{ \mathcal
J}^+(t)\rangle+C_-(t)|-\rangle\otimes|{\mathcal J}^-(t)\rangle$
where the bath states $|{\mathcal J}^{\pm}(t)\rangle$ are different.
The electron spin state is now given by the reduced density matrix
by tracing over the bath states, $\rho^s_{\sigma, \sigma'}(t) =
C^{\ast}_{\sigma'}(t) C_{\sigma}(t) \langle {\mathcal J}^{\sigma'}
(t) | {\mathcal J}^{\sigma}(t) \rangle$, $\sigma, \sigma'=\pm$. The
environment-driven shifting between $\rho^s_{+,+}$ and $\rho^s_{-,
-}$ is longitudinal relaxation. Either off-diagonal element gives a
measure of the coherence of the spin. The longitudinal relaxation
contributes to the decoherence. When this contribution is removed,
the remaining decoherence is called pure dephasing. For applications
in quantum technology, the longitudinal relaxation time $T_1$ can be
made much longer than the processing duration $T_p$ by a choice of
system and of  the electron Zeeman splitting much larger than the
dominant excitation energies in the bath and the spin-bath coupling
strength \cite{Fujisawa_T1,1shot_r_Kouwenhoven,Finley_spin_memory}.
When the cause of spin flip is removed, the reduced Hamiltonian of
the whole system is in the form diagonal in the electron spin basis,
$\hat{H}=|+\rangle \langle + | \otimes \hat{H}^+ + |-\rangle \langle
-| \otimes \hat{H}^-$. The bath evolves under the Hamiltonians
$\hat{H}^{\pm}$ into separate states $|{\mathcal J}^{\pm}(t)\rangle
\equiv e^{-i \hat{H}^{\pm} t} | {\mathcal J}\rangle$ depending on
the electron basis states $|\pm\rangle$. Pure dephasing is then
measured by $\mathcal{L}^s_{+,-}(t) = |\langle {\mathcal J}| e^{i
\hat{H}^- t} e^{-i \hat{H}^+ t}|{\mathcal J}\rangle |$. The electron
spin coherence may be restored by exploiting the dependence of the
bath dynamics on the electron spin states to make  the bifurcated
bath pathways intersect at a later time, i.e., $|{\mathcal J}^+ (t)
\rangle = |{\mathcal J}^-(t) \rangle$, leading to disentanglement.

At temperature ($\sim 10$mK$-1$K) $\gg$ the nuclear Zeeman energy
$\omega_n$ ($\sim $ mK) $\gg$ nuclear spin interaction ($\sim$ nK),
the nuclear bath initially has no off-diagonal coherence and is
described by $\sum_{\mathcal J} P_{\mathcal J} |\mathcal{J}\rangle
\langle \mathcal{J}|$ where $|{\mathcal J}\rangle \equiv \bigotimes
_n |j_n\rangle$, $j_n$ is the quantum number for $\hat{J}^z_n$, the
component of the $n$th nuclear spin along $z$ (the magnetic field
direction), and $P_{\mathcal J}$ gives thermal distribution. The
essence of electron decoherence is contained in the consideration of
each pure bath state $|\mathcal{J}\rangle$ and later the ensemble
average over $\mathcal{J}$ is included. The transverse interaction
$\hat{J}^+_n\hat{J}^-_m$ between two bath spins creates the
pair-flip excitation, $|j_n \rangle |j_m \rangle \rightarrow |j_n +1
\rangle |j_m -1 \rangle$. We sort out all such elementary
excitations from the ``vacuum" state $|\mathcal{J}\rangle$ and
denote each by the flip-process of a {\it pseudospin} $1/2$ indexed
by $k$: $|\uparrow_k\rangle \rightarrow |\downarrow_k \rangle$,
characterized by the energy cost $\pm E_k + D_k$ and the transition
matrix element $\pm A_k + B_k$ depending on the electron
$|\pm\rangle$ state. $\pm E_k $ ($\sim \mathcal{A}/N$) is from the
longitudinal interaction of the form $\hat{S}^{z} \hat{J}^{z}_n$
between the electron spin $\hat{S}^{z}$ and each of the two nuclear
spins \cite{e-n coupling}. $A_k $ ($\sim \mathcal{A}^2 /N^2 \Omega$,
$\Omega$ being the electron Zeeman energy) is the extrinsic coupling
between two nuclear spins mediated by virtual spin-flips of the
single electron \cite{Espin_YLS}. Its dependence on the number of
particles in the bath signifies its mesoscopic nature. $B_k$ is due
to the transverse part of the intrinsic (e.g. dipolar) interaction
between the nuclear spins with strength $\sim b$. The extrinsic
interaction $A_k$ couples any two spins in the meso-bath, as opposed
to the finite-range intrinsic interaction $B_k$. $D_k$ ($\sim b$) is
due to the longitudinal part of the intrinsic nuclear interaction.

In the nuclear bath with the descending order of parameters, $\Omega
\gg \omega_n \gg \mathcal{A}/N \gg b$, the bath dynamics is slow and
the density of pair-flip excitations created from the ``vacuum"
state $|\mathcal{J}\rangle$ is much less than unity in timescale of
interest \cite{Espin_YLS}. The excitations are almost always
spatially separated, leading to the pair-correlation approximation
\cite{Espin_YLS,Espin_HF_3_DasSarma} which treats pair-flips as
independent of each other. The bath is then driven by the effective
Hamiltonian derived from the first-principles interactions
\cite{Espin_YLS},
\begin{equation}
\hat{H}^\pm = \sum_k \hat{\mathcal H}^{\pm}_k\equiv \sum_k
{\mathbf h}_k^{\pm} \cdot \hat{\boldsymbol \sigma}_k/2,
\label{pseudospin_Hamil}
\end{equation}
where $\hat{\boldsymbol \sigma}_k$ is the Pauli matrix for
pseudospin $k$ driven by a pseudo-magnetic field ${\bf h}^{\pm}_k
\equiv (2B_k \pm 2 A_k, 0, D_k \pm E_k)$ depending on the electron
$|\pm\rangle$ state.

From the justification that correlations of more than two spins are
negligible \cite{Espin_YLS}, we derive the restrictions which the
decoherence timescale places on the size of the bath $N$, given by
$N^2 b^2 \mathcal{A}^{-2} \ll 1 \ll \min \left(\sqrt{N}, N^4 b^2
\Omega^4 \mathcal{A}^{-6} \right)$, to be established below. The
upper bound for $N$ distinguishes the bath from a macroscopic
system. It comes from the dominance of the pair correlation in the
interaction dynamics of the bath spins over the correlations of more
than two particles due to the intrinsic interaction. The lower
bound, $N^4 b^2 \Omega^4 \mathcal{A}^{-6} \gg 1$, is by a similar
consideration but due to the extrinsic interaction of the bath
spins. The lower bound $\sqrt{N}\gg 1$ simply signifies the
necessary statistics for decoherence. In the case of the electron
spin in a GaAs quantum dot, the theory is well justified for $10^8
\geq N \geq 10^4$ which covers quantum dots of all practical sizes.

The theory of the interacting nuclear spin dynamics dominated by the
pair excitation in the form of the pseudospin evolution leads to a
simple physical picture of coherence decay and restoration. The
initial unpolarized bath state $|\mathcal{J}\rangle = \bigotimes _n
|j_n\rangle$ can be replaced by the pseudospin product state
$\bigotimes_k|\uparrow_k\rangle$. Each pseudospin, representing a
nuclear spin states pair, initially points along the pseudospin
$+z$-axis and then precesses about the pseudo-magnetic field
$\mathbf{h}^{\pm}_k$, $|\psi_k^{\pm} \rangle \equiv
e^{-\frac{i}{2}{\mathbf h}^{\pm}_k\cdot\hat{\boldsymbol \sigma}_k t}
|\uparrow_k\rangle$, depending on the electron $|\pm \rangle$ state.
Thus, the electron spin coherence is measured by the divergence of
the pseudospin paths, $\mathcal{L}^s_{+,-}(t) = \prod_k |\langle
\psi_k^-| \psi_k^+ \rangle| \cong e^{-\sum_k \delta_k^2/2}$.
$\delta_k \equiv \sqrt{1-|\langle \psi_k^-| \psi_k^+ \rangle|^2}$ is
the geometric distance between the two conjugate pseudospin paths on
Bloch sphere.

\begin{figure}[t]
\includegraphics[width=8cm, height=5cm,bb=57 392 528 680,
clip=true]{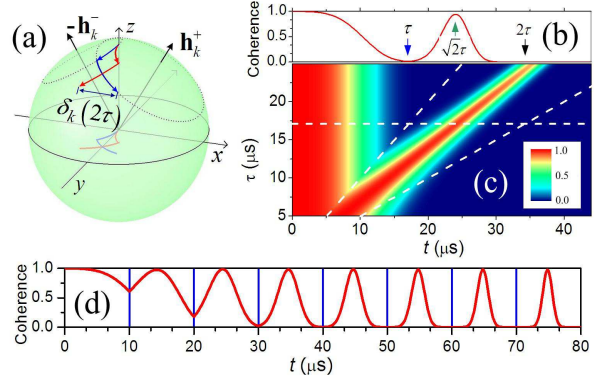} \caption{(a) Precession of the conjugated
pseudo-spin vectors under single-pulse control. Red (blue)
trajectory denotes when the electron state is $|+\rangle$
($|-\rangle$). (b) Electron spin coherence $\mathcal{L}^s_{+,-}(t)$
under the control of a single flip pulse applied at $\tau = 17~\mu$s
(indicated with the blue arrow). The revived coherence peaks at
$\sqrt{2}\tau$ (indicated by the green arrow) while coherence at the
conventional spin echo time $2\tau$ (indicated with the black arrow)
is negligible. (c) Contour plot of the electron spin coherence
$\mathcal{L}^s_{+,-}$ under single pulse control as a function of
time $t$ and the pulse delay time $\tau$ (indicated by the left
tilted line). The restoration of the coherence is pronounced at
$\sqrt{2}\tau$ whereas no coherence peak is visible at the
conventional echo time $2\tau$ (indicated by the right tilted line).
The horizontal line is the cut for the curve in (c). (d) Electron
spin coherence with a sequence of $\pi$-rotations (indicated by blue
vertical lines) at intervals of $\tau = 10~\mu$s. The numerical
evaluation is performed on a GaAs dot with thickness $d = 8.5$ nm in
growth direction [001] and lateral Fock-Darwin radius $r_0 = 25$ nm,
on the large-$N$ side of the mesoscopic regime where the intrinsic
nuclear interaction dominates \cite{Espin_YLS}. ${\bf B}_{\rm ext} =
10 ~$T
 along the [110] direction. The electron $g$-factor is
$-0.44$. The random choice of the initial bath state
$|\mathcal{J}\rangle \equiv \bigotimes_n |j_n\rangle$  from a
thermal ensemble at temperature $T=1$ K makes negligible difference
to the results.} \label{fig1}
\end{figure}

Now we examine the consequences of the pseudospin echo.
A fast $\pi$-pulse applied at $t=\tau$ to flip the electron spin \cite{Lyon_echo} would cause the
pseudo-spin evolution
\begin{equation}
|\psi^{\pm}_k(t)\rangle = e^{-\frac{i}{2}{\mathbf h}^{\mp}_k\cdot
\hat{\boldsymbol \sigma}_k(t-\tau)} e^{-\frac{i}{2}{\mathbf
h}^{\pm}_k\cdot \hat{\boldsymbol \sigma}_k\tau}|\uparrow\rangle.
\label{eq-pair}
\end{equation}
To find out how to control the decoherence, we neglect for the time
being the diagonal nuclear spin interaction $D_k$, which contributes
to the same component of the pseudo-magnetic field as $E_k$ but much
smaller. Because the pseudo-fields dominated by the extrinsic
nuclear spin interaction, $\mathbf{h}^{\pm}_k \equiv \pm (2A_k, 0,
E_k)$, invert exactly into each other, disentanglement of the
electron spin from the affected bath spin pairs follows at $2 \tau$
as in the classic spin echo to remove the inhomogeneous broadening
effect. The pseudo-fields dominated by the intrinsic interaction,
$\mathbf{h}^{\pm}_k \equiv (2B_k,0, \pm E_k)$, do not exactly invert
under the influence of the electron spin flip and the resultant
pseudospin trajectories are illustrated in Fig.~\ref{fig1}(a).
Disentanglement from the affected nuclear pairs and, hence, recovery
of the electron spin coherence occurs, by the rotation kinematics,
at time $\sqrt{2} \tau$, distinct from the classic echo.
Fig.~\ref{fig1}(b,c) give the computed results for  the electron
spin coherence in a dot of $10^6$ nuclear spins in GaAs which
reveals the coherence recovery after a flip of the electron spin at
a range of values for $\tau$, even after the coherence has
apparently vanished.

Furthermore, the coherence may be restored by a sequence of electron
spin flips. For example, with a sequence of $\pi$-pulses evenly
spaced with interval $\tau$, the disentanglement from the bath will
occur at $\sqrt{n(n+1)}\, \tau$ between the $n$th and the $(n+1)$th
pulses, as illustrated in Fig.~\ref{fig1}(d). Consider the
correction from the small term $D_k$ in the pseudo-fields, the
residue decoherence, at the disentanglement point
$\sqrt{n(n+1)}\,\tau$, is measured by $\delta^2_k \sim (E_k B_k D_k
\tau^3)^2$. Compared to the free-induction decay \cite{Espin_YLS}
where $\delta^2_k(\tau) \sim E^2_k B^2_k \tau^4$, the decoherence is
reduced by a factor of $\sim D^2_k \tau^2$ ($\sim 10^{-4}$ for $\tau
\sim 10 \mu$s).

Ensemble average over the mixed bath states is necessary in two
scenarios, namely, observation of decoherence of an ensemble of
quantum objects and observation of a single quantum object repeated
in a time
sequence~\cite{Lyon_echo,Marcus_T2,Koppens_Rabi,Gammon_t2star}. The
coherence of the electron spin is now $\rho_{+,-}(t) =C^{\ast}_- C_+
\mathcal{L}^s_{+,-}(t) \times \mathcal{L}^0_{+,-}(t)$, where
$\mathcal{L}^0_{+,-}(t) \equiv \sum_{\mathcal{J} } P_{\mathcal J}
e^{-i \phi_{\mathcal J} (t)}$ is the inhomogeneous broadening factor
due to the probability distribution $P_{\mathcal J}$ of the initial
bath state $|\mathcal{J}\rangle$ (different nuclear bath state may
result in different Overhauser energy-splitting ${\cal E}_{\mathcal
J}$ of the electron) \cite{Espin_YLS}. $\phi_{\mathcal J}(t) = {\cal
E}_{\mathcal
J}\left[\tau_1-(\tau_2-\tau_1)+\cdots+(-1)^n(t-\tau_n)\right]$ under
the control of a sequence of $\pi$-rotations on electron spin at
$\tau_1, \tau_2, \ldots,$ and $\tau_n$. The coherence factor
$\mathcal{L}^s_{+,-}(t)$ is insensitive, up to a factor of
$1/\sqrt{N} \ll 1$, to the selection of initial bath state
$|\mathcal{J}\rangle \equiv \bigotimes_n |j_n\rangle$ (verified by
numerical evaluations), and is taken out of the summation
\cite{Espin_YLS}.

\begin{figure}[t]
\includegraphics[width=8cm, height=6cm, bb=28 301 592 771, clip=true
]{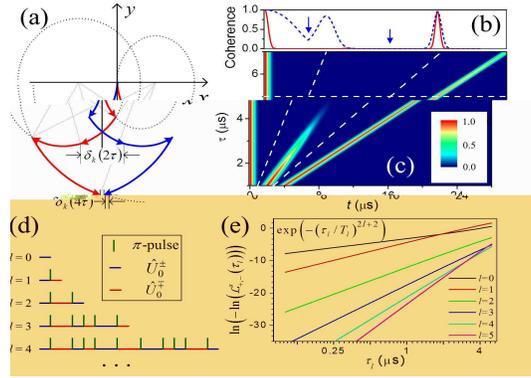} \caption{(a) Pseudo-spin trajectories (projected on
$x$-$y$ plane) driven by intrinsic nuclear interaction with two
$\pi$-rotations of the electron spin at $\tau$ and $3\tau$. (b)
Evolution of the electron spin coherence under two-pulse control
with $\tau=5~\mu$s. The dashed blue line denotes the pure state
dynamics part $\mathcal{L}^s_{+,-}$ and the solid red line includes
the inhomogeneous broadening factor  $ \mathcal{L}^0_{+,-}$. The
blue arrows indicate the times of the electron spin flip. (c)
Contour plot of the ensemble-averaged electron spin coherence under
the two-pulse control. The tilted lines indicate the pulse times and
the horizontal line is the cut for the curve in (b). In (b) and (c),
we have artificially set the ensemble dephasing time $T^{\ast}_2$ in
$\mathcal{L}^0_{+,-}$ to $0.5~\mu$s, about $100$ times greater than
its realistic value, to make the echo visible in the plot. (d)
Concatenated pulse sequences. The $(l+1)$th order sequence is
constructed by two subsequent $l$th order sequences, with a pulse
inserted if $l$ is even. (e) Dependence of echo magnitude on the
echo delay time $\tau_l$ under the control of the concatenated pulse
sequences in ensemble measurement. The quantum dot is  the same as
in Fig.~\ref{fig1} except smaller, $d = 2.8$ nm and $r_0 = 15$ nm.
The nuclear bath is assumed initially in thermal equilibrium at $T =
1~$K. } \label{fig2}
\end{figure}

Both the inhomogeneous broadening and the pure decoherence due to
the extrinsic bath spin interaction are shown to be removed at the
classic spin echo time $2\tau$ in contrast to the unusual recovery
time of $\sqrt{2}\tau$ in the case of intrinsic bath spin
interaction. We need a pulse sequence to produce a time where the
decoherence from all three sources can be removed. A solution is a
two-pulse control. Fig.~\ref{fig2}(a) shows that, after a second
electron spin flip at $3\tau$,  the paths of the two pseudospin
states corresponding to the electron $|\pm\rangle$ states driven
by the intrinsic nuclear spin pair interaction cross again at
$4\tau$, coinciding with the secondary spin echo time for the
other two causes. This two pulse sequence is well-known as
Carr-Purcell sequence in NMR spectroscopies \cite{Slichter_NMR}.
The residual decoherence at $t=4 \tau$ is $\delta^2_k \sim
16\left(E_kB_k-A_kD_k\right)^2D_k^2\tau^6$. The restoration by
two-pulse control of coherence in the presence of both pure and
ensemble decoherence  is demonstrated by the results of numerical
evaluation shown in Fig.~\ref{fig2}(b,c). The electron spin
coherence is restored at $4 \tau$ by the second pulse even when
the first spin-echo at $2\tau$ has completely vanished,
illustrating the remarkable observation \cite{Pines_spinecho} that
the absence of spin echo does not mean irreversible loss of
coherence.

The power of concatenation of pulse sequences has been shown in the
context of dynamical decoupling in quantum computation
\cite{concatenation_Lidar}. Similarly, the control of
disentanglement of the bath states from the quantum object may be
enhanced by concatenation. The pseudo-spin evolution with the
two-pulse control of the quantum object can be constructed
recursively from the free-induction evolution
$\hat{U}^{\pm}_0=e^{-i\mathbf{h}^{\pm}_k\cdot\hat{\boldsymbol\sigma}_k\tau/2}$,
by the concatenation, $\hat{U}^{\pm}_l =
\hat{U}^{\mp}_{l-1}\hat{U}^{\pm}_{l-1}$, $l=1,2$.  The process can
be extended by iteration, Fig.~\ref{fig2}(d), to any level,
$\hat{U}^{\pm}_l =
e^{-i{\boldsymbol\theta}^{\pm}_{l}\cdot\hat{\boldsymbol{\sigma}}_k/2}$,
where ${\boldsymbol\theta}^{\pm}_{l}$ is the rotation vector along
the axis of rotation  through an angle $\theta^{\pm}_{l}$.
Disentanglement occurs at $\tau_l\equiv2^l \tau$ coinciding with the
classic spin echo. For small $\theta^{\pm}_{l}$, the recursion
relation is
${\boldsymbol\theta}^{\pm}_{l+1}={\boldsymbol\theta}^{+}_{l}+{\boldsymbol\theta}^{-}_{l}
\mp {\boldsymbol\theta}^{+}_{l}\times {\boldsymbol\theta}^{-}_{l} $.
At each iteration, the rotation vectors of the conjugate pseudo-spin
states have their mean
$\left({\boldsymbol\theta}^{+}_{l}+{\boldsymbol\theta}^{-}_{l}\right)/2$
increased by a factor of $2$ and their difference
$\left({\boldsymbol\theta}^{+}_{l}-{\boldsymbol\theta}^{-}_{l}\right)$
reduced by a factor of ${\theta}^{\pm}_{l}\sim 2^l b \tau$ (deduced
by induction from $\theta^{\pm}_1 \sim  2 b \tau$). The decoherence
is reduced by an order of $b^2\tau_l^2$ at $\tau_l$ for each
additional level of concatenation till saturation at the level
$l_0\approx -\log_2(b \tau)$. Hence, the coherence echo magnitude
scales with the echo delay time according to $\exp\left(-(\tau_l
/T_l)^{2l+2} \right)$ as shown in Fig.~\ref{fig2}(e). Our result
shows the protection of electron spin coherence  by pulse sequences
with interpulse interval up to $\sim 10 \mu$s

In conclusion, we note that our scheme of restoring the coherence
depends on the pure decoherence being driven by the interaction in
the spin bath and by the domination of the bath pair excitation in
the slow bath dynamics. The pulse sequence design is borrowed from
the dynamical decoupling schemes in NMR spectroscopies
\cite{Slichter_NMR} and in quantum computation
\cite{concatenation_Lidar} but the disentanglement method aims
directly at the bath dynamics. Our method seeks not to eliminate the
object-bath interaction by dynamical averaging, but to disentangle
by controlling the quantum object to maneuver the bath evolution.
Thus, elimination of coupling between the quantum object and the
bath is not a necessary condition for their disentanglement, as
illustrated by coherence recovery at $t=\sqrt{n(n+1)}\tau$ where
effective object-bath interaction does not vanish even in the first
order of hyperfine coupling $\mathcal{A}/N$. Direct observation of
coherence echoes at such magic times is possible with the narrowing
of inhomogeneous distribution by measurement projection
\cite{inhom_filter}. The control of bath spins may well develop into
a valuable addition to the collection of armaments of coherence
preservation for quantum information processing.

We acknowledge support from NSF DMR-0403465, ARO/LPS, and
DARPA/AFOSR.

\pagebreak

\begin{center}
{\large \textbf{Supplementary Information for ``Restoring Coherence
Lost to a Slow Interacting Mesoscopic Bath''}}
\end{center}

\section*{Pseudo-spin model}

We give the salient points of the method of solving the electron
nuclear spin dynamics in a GaAs quantum dot. The system consists of
an electron with spin vector $\hat{{\mathbf S}}_{\rm e}$ and $N$
nuclear spins, $\hat{{\mathbf J}}_{n}$, with Zeeman energies
$\Omega$ and $\omega_{n}$ under a magnetic field $B_{\rm ext}$,
respectively, where $n$ denotes both positions and isotope types
($^{75}$As, $^{69}$Ga and $^{71}$Ga). The interaction can be
separated as ``diagonal'' (or ``longitudinal'')  terms which involve
only the spin vector components along the field ($z$) direction and
``off-diagonal'' (or ``transverse'') terms which involve spin flips.
Because $\Omega$ is much larger than $\omega_n$ and the strength of
the direct nuclear-nuclear interactions (e.g., nuclear dipolar
interaction), the off-diagonal part of the electron nuclear
hyperfine interaction can be eliminated by a standard canonical
transformation, with the second-order correction left as the
hyperfine-mediated nuclear-nuclear interaction. For the same reason,
the off-diagonal part of the nuclear-nuclear interaction includes
only terms which conserve the Zeeman energies, thus excluding the
hetero-nuclear terms. The total effective Hamiltonian can be written
as $\hat{H}=\hat{H}_{\rm e}+\hat{H}_{\rm N}+\sum_{\pm}|\pm\rangle
\hat{H}^{\pm} \langle\pm|$, with $\hat{H}_{\rm e}=\Omega
\hat{S}^z_{\rm e}$, $\hat{H}_{\rm N}=\omega_{n}\hat{J}^z_{n}$,
$\hat{H}^{\pm}=\pm \hat{H}_A+\hat{H}_B+\hat{H}_D\pm \hat{H}_E$, and
\begin{subequations} \begin{eqnarray}
\hat{H}_A &=&
  {\sum_{n\ne m}}'\frac{a_n a_m}{4  \Omega}
  \hat{J}_n^{+}\hat{J}_m^{-}\equiv {\sum_{n \ne m}}' A_{n,m} \hat{J}_n^{+}\hat{J}_m^{-}, \ \ \ \   \label{HA} \\
\hat{H}_B &=&{\sum_{n \ne  m}}' B_{n,m} \hat{J}_n^{+}\hat{J}_m^{-}
\label{HB}  \\
\hat{H}_D &=&{\sum_{n < m}} D_{n,m} \hat{J}_n^{z}\hat{J}_m^{z}
\label{HD}  \\
\hat{H}_E &=&
 \sum_{n}\left(a_n/2\right)\hat{J}_n^{z}\equiv \sum_{n}E_n \hat{J}_n^{z},   \label{HE}
\end{eqnarray} \label{Hamiltonian}
\end{subequations}
where $|\pm\rangle$ are the eigenstates of $\hat{S}^z_{\rm e}$, the
summation with a prime runs over only the homo-nuclear pairs, the
subscript $A$ denotes the hyperfine mediated nuclear-nuclear
interaction, $B$ the off-diagonal part of the direct nuclear-nuclear
interaction, $D$ the diagonal part of the direct nuclear-nuclear
interaction, and $E$ the diagonal part of the contact
electron-nuclear hyperfine interaction. The hyperfine energy,
determined by the electron orbital wavefunction, has a typical
energy scale $E_n\sim a_n \sim \frac{\mathcal{A}}{N} \sim
10^6$~s$^{-1}$ for a dot with about $10^6$ nuclei~\cite{Paget},
where $\mathcal{A}$ is the hyperfine constant depending only on the
element type. The direct nuclear-nuclear interaction, which is
``short-ranged" (referred here as decaying no slower than dipolar),
has the near-neighbor coupling $B_{n,m}\sim {D}_{n,m} \sim b \sim
10^2$~s$^{-1}$. The hyperfine mediated interaction, which couples
any two nuclear spins that are in contact with the electron and is
associated with opposite signs for opposite electron spin states,
has an energy scale dependent on the field strength, $A_{n,m}\sim
\frac{\mathcal{A}^2}{N^2 \Omega} 1$--$10$~s$^{-1}$ for field $\sim
40$--$1$~T. This hyperfine mediate interaction is differentiated
from the ``short-range" direct nuclear-nuclear interaction by the
qualifier ``infinite-range". We work in the interaction picture
defined by $\hat{H}_e$ and $\hat{H}_N$ in which the dynamics are
determined by $\hat{H}^{\pm}$.

\begin{figure}[b]
\includegraphics[width=17cm, height=6cm,bb=54 552 563 731,
clip=true]{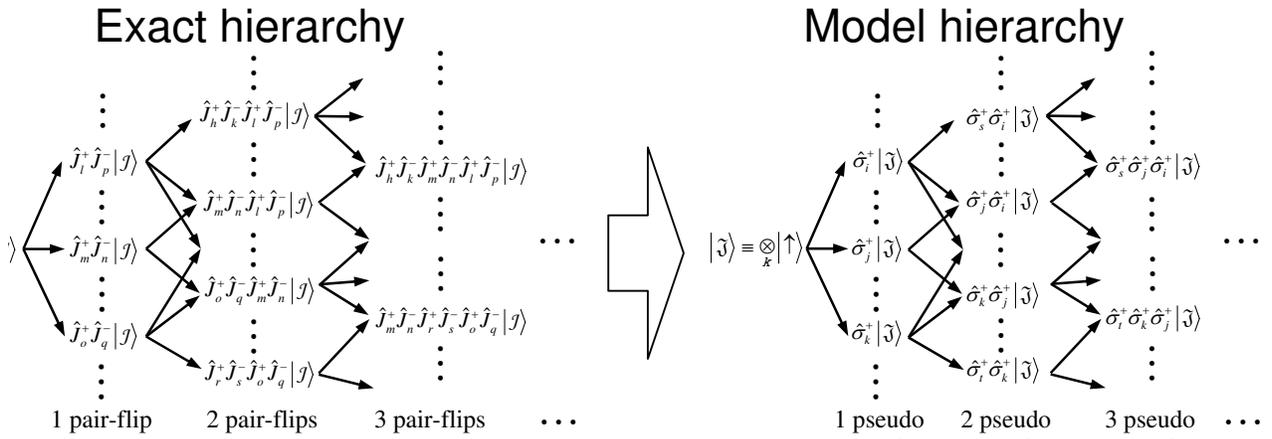} \caption{Hierarchy of the nuclear spin
dynamics.} \label{hierarchy}
\end{figure}

The basis set of the bath are eigenstates of $\hat{H}_N$:
$\bigotimes_n |j_n\rangle$. In Eqn.~(\ref{Hamiltonian}), $\hat{H}_D$
and $\hat{H}_E$ are diagonal in this basis. The off-diagonal terms
$\hat{H}_A$, $\hat{H}_B$ are weak perturbations that will excite the
bath initially on an arbitrary configuration $| \mathcal{J} \rangle
\equiv |j_1\rangle \cdots |j_N \rangle$. The elementary excitations
in the nuclear spin bath are pair-flip excitations created by
operators $\hat{J}_m^+\hat{J}_n^-$ in the reduced Hamiltonian.
Starting from any initial nuclear configuration, the evolution of
the nuclear spin states by these elementary excitations is of the
hierarchy as shown in the left side of Fig.~\ref{hierarchy}. We can
regard the zeroth layer of this hierarchy, the initial state, as the
`vacuum' of the pair-flip excitations and layer $n$ corresponds to
$n$ pair-flip excitations have been created. The state at time $t$
is a linear superposition of all possibilities:
\begin{equation}
|\mathcal{J}(t)\rangle = C_{\mathcal J}(t) |\mathcal{J}\rangle +
\sum_{m,n} C_{m,n}(t)\hat{J}_m^+\hat{J} _n^- |\mathcal{J}\rangle +
\sum_{l,p,m,n} C_{l,p,m,n}(t)\hat{J}_l^+\hat{J}_p^-
\hat{J}_m^+\hat{J}_n^- |\mathcal{J}\rangle + \cdots.
\label{expansion_exact}
\end{equation}
where the summation over the indexes $m,n,l,p,\ldots$ are defined
such that $|\mathcal{J}\rangle$, $\hat{J}_m^+\hat{J} _n^-
|\mathcal{J}\rangle$, $\hat{J}_l^+\hat{J}_p^- \hat{J}_m^+\hat{J}_n^-
|\mathcal{J}\rangle$, $\ldots$ denote different eigenstates of
$\hat{H}_N$ orthogonal to each other.

\begin{figure}[p]
\includegraphics[width=16cm, height=18.5cm,bb=81 200 543 710,
clip=true]{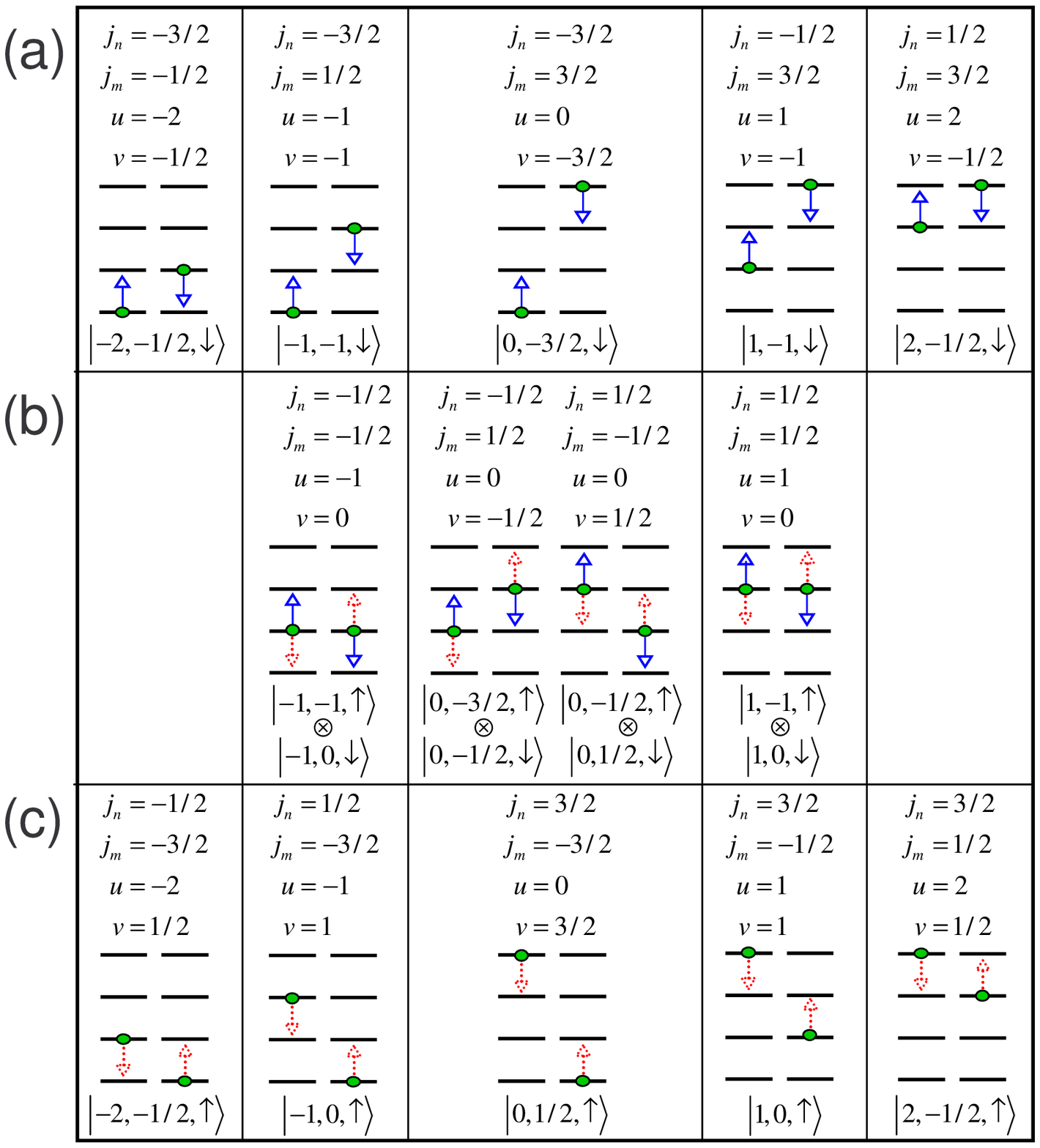} \caption{Illustration of construction of
pseudo-spin states from a pair of nuclei of spin 3/2. The solid dots
show state $|j_n\rangle |j_m\rangle$. The solid arrows between lines
show to which state the operator $\hat{J}_n^+\hat{J}_m^-$ would lead
and the dotted arrows show to which state $\hat{J}_n^-\hat{J}_m^+$
would lead. (a)Monogamy states which are mapped to pseudo-spin down:
$|u, v, \downarrow \rangle$. (b)Bigamy states which are mapped to
two pseudo-spins with one up and one down respectively: $|u, v-1,
\uparrow \rangle \otimes |u, v, \downarrow \rangle$ (c) Monogamy
states mapped to pseudo-spins up: $|u, v-1, \uparrow \rangle$.}
\label{map}
\end{figure}

We solve this dynamics in the nuclear spin bath based on a
pseudo-spin model as described below. We have, on sites labelled
$n=1, \ldots,N$, the nuclear spin states $|j_n\rangle$ with $ -j
\leq j_n \leq j$ for nuclei of spin $j$. As the elementary
excitations are pair dynamics driven by $\hat{J}^+_m \hat{J}^-_n$,
we first sort out the pair states $|j_n\rangle |j_m \rangle$. These
pair states are divided into three categories:

1. {\it Down States}: A down state $|j_n\rangle |j_m \rangle$ has a
partner $| j_n+1\rangle |j_m-1 \rangle$ created by
\[
\hat{J}^+_n \hat{J}^-_m|j_n\rangle |j_m\rangle = \sqrt{(j + j_n
+1)(j-j_n)} \sqrt{(j - j_m +1)(j+j_m)} |j_n+1\rangle |j_m-1\rangle
\]
A down state must have ($j_n < j, j_m > -j$). There are $(2j)^2$
down states for each bond.

2. {\it Up States}: An up state $|j_n\rangle |j_m \rangle$ has a
partner $| j_n-1\rangle |j_m+1 \rangle$ created by $\hat{J}^-_n
\hat{J}^+_m|j_n\rangle |j_m\rangle$. The up state must have ($j_n
> -j, j_m < j$). There are $(2j)^2$ up states for each bond.

3. {\it Bachelor States}: A single pair state has no partners
connected by $\hat{J}^-_n \hat{J}^+_m|j_n\rangle |j_m\rangle$ or its
Hermitian conjugate, i.e., $j_n=j_m=j$ or $j_n=j_m=-j$.

The Bachelor states may be mapped to pseudo-spin 0 states. Since
they are scalar states, their Hamiltonian terms will commute with
every other operators, they can only contribute to the phase factor
in the electron spin coherence through the Overhauser field, causing
inhomogeneous broadening.

We shall find by an explicit construction that the up and down
states can be paired to provide states of $(2j)^2$ two-level systems
- the pseudo-spins. These states are divided into:

1. {\it Monogamy States}: Each state belongs to only one pseudo-spin
although its partner may be a bigamist. These states are edge states
in that at least one of the two spin quantum numbers ($j_n$ or
$j_m$) equal to $\pm j$ but they cannot both be equal to $j$ or to
$-j$. Half of them ($4j-1$ states) are down states, $|j_n=-j \rangle
|j_m>-j \rangle$ or $|j_n<j\rangle |j_m=j \rangle$, see
Fig.~\ref{map}(a). The other $4j-1$ states, $|j_n>-j\rangle
|j_m=-j\rangle$ or $|j_n=j\rangle |j_m<j \rangle$ are up states, see
Fig.~\ref{map}(c).

2. {\it Bigamy States}: Each belongs to two different pseudo-spins.
They are interior states: $-j <j_n< j$ and $-j <j_m< j$. There are
$(2j-1)^2$ of them, see Fig.~\ref{map}(b).

For a bond between two sites $(n,m)$, all possible pseudo-spins
excitations are sort out and labelled by using two numbers $u(j_n,
j_m)$ and $v(j_n, j_m)$ for $-j \leq j_n, j_m \leq j$,
\begin{eqnarray}
u &=& j_m + j_n \\
v &=& \frac{1}{2} (j_n - j_m) \notag
\end{eqnarray}
The construction follows from the angular momentum addition of the
two sites and is illustrated for $j=3/2$,
\begin{eqnarray}
\begin{array}{crrrrrrr}
u = &  -3 & -2  & -1 & 0 & 1 & 2 & 3    \\
   &   &   &   & \frac{3}{2} &   &  &   \\
 &     &   & 1 &  & 1 &  &    \\
   &     & \frac{1}{2}  & & \frac{1}{2} &   & \frac{1}{2}   &   \\
v = &   0 &  & 0 &  & 0 &  & 0 \\
   &     &- \frac{1}{2}  & & -\frac{1}{2} &   & -\frac{1}{2}   &   \\
&     &   & -1 &  & -1 &  &    \\
   &   &   &   &- \frac{3}{2} &   &  &
\end{array}
\end{eqnarray}
When $u = \pm 2j$, the two-spin states are
\begin{eqnarray}
|j_n \rangle |j_m\rangle = |j\rangle |j\rangle ~~~~\text{or} ~~~~
|-j \rangle |-j\rangle.
\end{eqnarray}
Both are Bachelor states. For a $u$ value equal to or between
$-2j+1$ and $2j-1$, the $v$ state at the bottom of the ladder is
chosen to be the spin down of the doublet and one rung above to be
its partner. Thus, the bottoms of the ladders yield the down
monogamy states and the top rungs the up monogamy states. In
between, the states are the bigamy states.

For any initial configuration $|\mathcal{J}\rangle$, the relevant
set of pseudo-spins $ \mathcal{G}_{\mathcal J}$ is determined by
examining every possible nuclear spin pair $ (m,n)$. Each pair will
contribute $0$, $1$ or $2$ pseudo-spins if $|j_m\rangle |j_n
\rangle$ is in the Bachelor state, Monogamy state or Bigamy state
configuration respectively (see Fig.~\ref{map}). The many nuclear
spin initial state $ |\mathcal{J}\rangle$ is then replaced by,
\begin{equation}
|\mathcal{J}\rangle \equiv |j_1\rangle \cdots |j_N \rangle
\Rightarrow | \mathfrak{J}\rangle \equiv \bigotimes_{k \in
\mathcal{G}_{\mathcal J} } |k \sigma \rangle \label{mapping_state}
\end{equation}
where $k$ labels both the nuclear pair $(m,n)$ and the pseudo-spin
type $(u, v)$. $\sigma = \uparrow$ or $\downarrow$ depending on
whether $k$ is mapped from up or down Monogamy or Bigamy state.
Different initial nuclear configurations will result in different
sets of pseudo-spins. For a randomly chosen initial configuration
$|\mathcal{J}\rangle$, the number of pseudo-spins is given by $M
\sim (\frac{2j}{2j+1})^2ZN$ where $N$ is the total number of nuclear
spins and $Z$ the number of nuclei coupled to a particular nuclear
spin by the nuclear-nuclear interaction. For short ranged direct
interaction, $Z \sim {\rm O}(10)$ and for the infinite-ranged
hyperfine mediated interaction $Z = N$. The factor
$(\frac{2j}{2j+1})^2$ arises as the single state, Monogamy state and
Bigamy state are contributing $0$, $1$ and $2$ pseudo-spins
respectively. For convenience, when the set $\mathcal{G}_{\mathcal
J}$ is determined from the $|\mathcal{J}\rangle$, we redefine the
pseudo-spin up and down states, i.e. $|k,\sigma\rangle \rightarrow
|k, -\sigma\rangle$, for those pseudo-spins in set
$\mathcal{G}_{\mathcal J}$ so that the initial state $|\mathfrak{J}
\rangle$ in this new definition corresponds to all pseudo-spins
pointing `up': $\bigotimes_k |\uparrow\rangle_k$. The redefinition
is conditioned on the initial nuclear configuration
$|\mathcal{J}\rangle$.

\begin{figure}[t]
\includegraphics[width=9.8cm, height=6cm,bb=61 327 555 662,
clip=true]{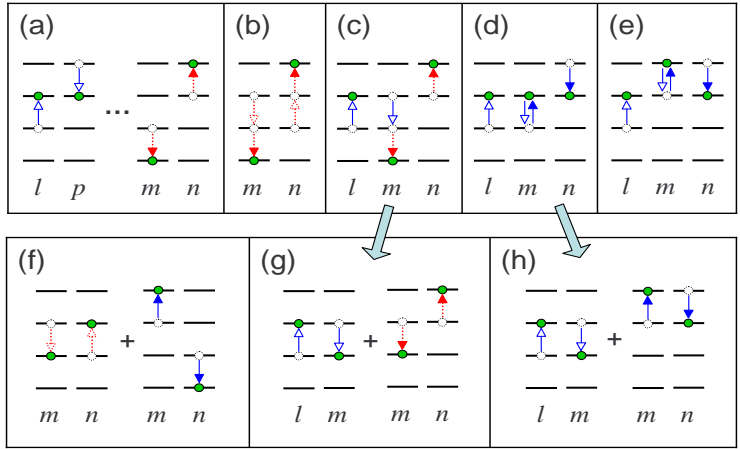} \caption{Illustration of multi pair-flip
excitations in the nuclear bath. We use hollow arrowheads to
indicate the first pair-flip and solid arrowheads to indicate the
second pair-flip. Dotted arrows denote the pseudo-spin flip from
'down' state to 'up' state and solid arrows for the inverse process
(see text). (a) Independent pair flips; (b-e) Various situations of
overlapping pair-flips; (f-h) Approximations in the independent
pseudo-spin model.} \label{ovelap}
\end{figure}

Conditioned on the electron state $|\pm\rangle$, the pseudo-spins
are driven by the effective Hamiltonian of the form,
\begin{equation}
\hat{H}^{\pm}_{\rm sp} = \sum _k \hat{\mathcal{H}}_k^{\pm} \equiv
\sum_{k} {\bf h}^{\pm}_{k} \cdot \hat{\boldsymbol \sigma}_{k}/2
\label{mapping_hamiltonian}
\end{equation}
The effective magnetic field ${\bf h}_k^{\pm}$ on the pseudo-spins,
conditioned on the electron spin state, are to be determined by
reproducing the matrix elements, $\langle \mathcal{J} | \hat{J}^+_m
\hat{J}^-_n \hat{H}^{\pm} \hat{J}^+_n \hat{J}^-_m | \mathcal{J}
\rangle - \langle \mathcal{J} | \hat{H}^{\pm} | \mathcal{J} \rangle
$ and $\langle \mathcal{J} | \hat{H}^{\pm} \hat{J}^+_n \hat{J}^-_m |
\mathcal{J} \rangle $, namely the energy cost and transition matrix
element for nuclear pair-flips.

The pseudo-spin model for characterizing the nuclear spin bath
dynamics is to approximate the exact evolution of
Eqn.~(\ref{expansion_exact}) by the independent evolution of all
pseudo-spins in $\mathcal{G}_{\mathcal J} $,
\begin{equation}
|\mathfrak{J}(t)\rangle = \bigotimes_{k \in \mathcal{G}_{\mathcal J}
} |\psi^{\pm}_k(t)\rangle = C_{\mathfrak{J}}(t)|\mathfrak{J}\rangle
+ \sum_{k_1} C_{k_1}(t) \hat{\sigma}^+_{k_1}|\mathfrak{J}\rangle +
\sum_{k_1,k_2} C_{k_1,k_2}(t) \hat{\sigma}^+_{k_1}
\hat{\sigma}^+_{k_2}|\mathfrak{J}\rangle + \cdots.
\label{expansion_model}
\end{equation}
This pseudo-spin dynamics can be put into a similar hierarchy as
shown in the right part of Fig.~\ref{hierarchy} which we will refer
to as model hierarchy in contrast to the exact hierarchy.

With the mapping established for the state
(Eqn.~(\ref{mapping_state})) and the Hamiltonian
(Eqn.~(\ref{mapping_hamiltonian})), the first two layers of the
exact hierarchy will be reproduced exactly by the model hierarchy,
i.e., there is a one to one correspondence between $\hat{J}_m^+
\hat{J}_n^-|\mathcal{J}\rangle$ and $\hat{\sigma}
_k^+|\mathfrak{J}\rangle$ with the energy and coupling to the
initial state $ | \mathcal{J} \rangle$ ($|\mathfrak{J}\rangle$)
exactly reproduced.

Difference between the model and the exact hierarchies arises when
more than one excitations have been created in the system. In
Fig.~\ref{ovelap}, we illustrate with the case when two pair-flip
excitations have been created. If the two pair-flips do not overlap
as shown in Fig.~\ref{ovelap}(a), their dynamics are then
independent of each other and well described by the pseudo-spin
model. Fig.\ref{ovelap}(b-e) illustrate the various situations that
the two pair-flips overlap, by sharing one or two nuclei. The
flip-flop of the first nuclear pair $(l,m)$ changes the spin
configuration of both nuclear $l$ and $m$ and if a second flip-flop
is to take place on pair $(l,m)$ or $(n,m)$ or $(l,n)$, it is no
longer described by the dynamics of the original pseudo-spins
assigned to it. Instead, in the model hierarchy by the independent
pseudo-spin model, two successive flip-flops on pair $(l,m)$ or two
successive flip-flops on pair $ (l,m)$ and $(m,n)$ respectively are
shown in Fig.\ref{ovelap}(f-h). Fig.~\ref {ovelap}(g) can be
considered as the approximate form of Fig.~\ref{ovelap}(c) and
Fig.~\ref{ovelap}(h) as that of Fig.~\ref{ovelap}(d). The model
hierarchy contains events like Fig.~\ref{ovelap}(f) which is absent
in the exact hierarchy and events like Fig.~\ref{ovelap}(e) in the
exact hierarchy is not contained in the model hierarchy. Therefore,
on layer $2$, the model hierarchy coincides with the exact hierarchy
in events described by Fig.~\ref{ovelap}(a) and differ by replacing
the events of Fig.\ref{ovelap}(b-e) with events of
Fig.\ref{ovelap}(f-h). The difference in a general layer can be
analyzed in the same way.

By the pseudo-spin model, we are using Eqn.~(\ref{expansion_model})
as the bath state at time $t$ for calculating physical properties
instead of Eqn.~(\ref{expansion_exact}). The difference of the exact
and model hierarchies is estimated below which serves as an upper
bound for error estimation (notice that the physical properties of
interest are not necessarily changed by replacing events of
Fig.\ref{ovelap}(b-e) with events of Fig.\ref{ovelap}(f-h),
therefore, this error-estimation is not necessarily a tight bound).
If $n-1$ pair-flip excitations have already been generated, to
create the next excitation, we have $M$ pseudo-spin to choose from
and $\sim 2(n-1)Z$ of them overlap with the previous excitations.
Therefore, the probability of having a new excitation without
overlapping with the previous excitations is given by: $\sim \frac{M-2(n-1)Z%
}{M}$. By induction, the probability of creating $n$ non-overlapping
pair-flip excitations is then given by,
\begin{eqnarray}
p(n) & \simeq& 1 \times \frac{M-2Z}{M} \cdots \times \frac{M -
2(n-1)Z}{M}
\label{errorprob} \\
& \simeq & \mathrm{exp} \left[ - \frac{2Z}{M} - \cdots - \frac{
2(n-1)Z}{M} \right]
\notag \\
&=&\mathrm{exp} \left[ -\frac{n(n-1)}{N}(\frac{2j+1}{2j})^2 \right]
\notag
\end{eqnarray}
The second $\simeq$ holds if $2nZ \ll M$ which is always true in the
timescale relevant in our study.

Comparing the model hierarchy and the exact hierarchy, we find from
the above analysis that they differ in layer $n$ with the relative
amount of $ 1-p(n)$.

\section*{Error estimation}

We perform a self-consistent analysis on validity of the pseudo-spin
model. Here the relevant timescale plays the crucial role in
determine the validity of the model. If only the first $n$ layers of
the hierarchy (the exact one and the model one) are involved, the
error in the calculated physical properties is bounded by $1-p(n)
\ll 1$ if $n^2 \ll N$. Therefore, in the very short time limit, only
the first several layers of the hierarchy can be involved and the
pseudo-spin model gives an almost exact account of the dynamics. To
estimate error upper bound for the longer time limit, we will
calculate the number of layers involved (denoted as $n$) based on
the model hierarchy of the pseudo-spin model. If $n^2/N$ obtained is
small, we conclude that $n$ also faithfully reflects the number of
layers involved in the exact hierarchy. Therefore, the error
estimation based on the pseudo-spin model is faithful and any
physical properties calculated based on pseudo-spin model is also a
good approximation since the difference from the exact dynamics is
small. Otherwise, the approximation is not good. It is established
below that the condition $n^2 \ll N$ is the origin of an upper bound
and a lower bound on the quantum dot size $N$ for which our theory
is able to deal with. The validity of the pair-correlation approach
(pseudo-spin model) is indeed in the mesoscopic regime.

At any given time $t$, an average excitation number $N_{flip}(t)$
can be defined as follows based on the model hierarchy,
\begin{equation}
N_{flip}(t)=1\times \sum_{k_{1}}\left| C_{k_{1}}(t)\right|
^{2}+2\times \sum_{k_{1},k_{2}}\left| C_{k_{1},k_{2}}(t)\right|
^{2}+3\times \sum_{k_{1},k_{2},k_{3}}(t)\left|
C_{k_{1},k_{2},k_{3}}\right| ^{2}+\cdots \label{nflip}
\end{equation}
and our analysis \cite{Espin_unpub_data} shows that the
layer-distribution of population in the hierarchy is of a normal
distribution centered at $N_{flip}$, i.e. the population is
distributed in layers from layer $N_{flip}- \sqrt{N_{flip}}$ to
$N_{flip} + \sqrt{N_{flip}}$. Therefore, the quantity for
characterizing the error upper bound is of a very simple form:
$P_{err}(t) \equiv 1-\mathrm{exp} (-N_{flip}^2(t) / N ) $.

In the pseudo-spin model, $N_{flip}(t)$ defined in
Eqn.~(\ref{nflip}) has an equivalent expression which is more
convenient for evaluation:
\begin{equation}
N_{flip}(t)=\sum_k |\langle \downarrow| U^{\pm}_k(t) | \uparrow
\rangle|^2
\end{equation}
where $U^{\pm}_k(t) \equiv e^{-i\hat{\mathcal{H}}^{\pm}_k t}$ is the
evolution operator for pseudo-spin $k$. The contribution can be
divide into two parts:$N_{flip}\left( t\right) =N_{flip}^{A}\left(
t\right) +N_{flip}^{B}\left( t\right) $. $N_{flip}^{A}$ is the
number of non-local pair-flip excitations and $N_{flip}^{B}$ is the
number of local pair-flip excitations that have been created. $
N_{flip}^{A}\left( t\right) $ and $N_{flip}^{B}\left( t\right) $
have very different behavior and we analyze them separately.

In free-induction evolution, the number of non-local pair-flip
excitations is given by,
\[
N_{flip}^{A}\left( t\right) =\sum_{k}\left( \frac{
2A_{k}}{h_{k}^{A}}\right) ^{2}\sin ^{2}\frac{h_{k}^{A}t }{2} \leq
\sum_{k} A_{k}^{2} t^2 \simeq M_A \frac{\mathcal{A}^4}{N^4\Omega^2}
t^2 \simeq \frac{\mathcal{A}^4}{N^2\Omega^2} t^2
\]
where $h_{k}^{A}\equiv \sqrt{ E_{k}^{2}+4A_{k}^{2}}$ and $M_A \sim
N^2$ is the number of non-local nuclear spin pairs. Since the
evolution of the non-local pair-correlation is completely reversed
by the $\pi$ pulses, $N_{flip}^{A}\left( t\right)$ is also reversed
and $N_{flip}^{A}=0$ at each spin echo time. Therefore,
$N_{flip}^{A}\left( t\right) $ does not accumulate in the pulse
controlled dynamics and we just need to look at the maximum value of
$N_{flip}^{A}\left( t\right) $ between echoes. For example, in the
control with the equally spaced pulse sequence, in order to have the
coherence well preserved or restored at spin echo time, the delay
time between successive pulses is limited by $\tau \lesssim T_{H} $,
where $T_H \simeq b^{-1/2} \mathcal{A}^{-1/2} N^{1/4}$ is the Hahn
echo decay timescale and $\sim 10\mu s$ in GaAs \cite{Espin_YLS}.
Therefore, $N_{flip}^{A}$ at any time is bounded by
$N_{flip}^{A}\left( {\rm O}(T_{H})\right)$ in this scenario. The
estimate of the bound on $N$ imposed by $N_{flip}^A$ for the
validity of our approach is therefore,
\begin{equation}
\left( \frac{\mathcal{A}^4}{N^2\Omega^2} T_{H}^2 \right)^2
\frac{1}{N} \sim \frac{\mathcal{A}^6}{N^4\Omega^4 b^{2}} \ll 1
\end{equation}
For GaAs fluctuation dot in a magnetic field of $10$ Tesla, the
above condition is well satisfied for $N \gtrsim 10^4$. Estimation
for other controlling pulse sequences give a lower bound on $N$
which is in the same order.

For the number of local pair-flip excitations, we have a similar
expression in the free-induction evolution,
\[
N_{flip}^{B}\left( t\right) =\sum_{k}\left( \frac{2B_{k}
}{h_{k}^{B}}\right) ^{2}\sin ^{2}\frac{h_{k}^{B}t}{2} \leq \sum_{k}
B_{k}^{2} t^2 \simeq M_B b^2 t^2 \simeq \alpha N b^2 t^2
\]
where $ h_{k}^{B}\equiv \sqrt{E_{k}^{2}+4B_{k}^{2}}$ and $M_B \sim
\alpha N$ is the number of local nuclear spin pairs. $\alpha$ ($
\sim 10$ in GaAs) is determined by number of local neighbors and the
nuclear spin quantum number $j$. In contrast to the non-local pair
dynamics, the local pair dynamics is {\it not} reversed under the
influence of the electron spin flip and $N^B_{flip}(t)$ accumulates
all through the time. Nonetheless, it turns out that $N^B_{flip}(t)
\leq \sum_{k} B_{k}^{2} t^2 \simeq \alpha N b^2 t^2$ holds for all
scenarios of pulse controls being discussed. Therefore, the
condition $(N^B_{flip})^2/N \ll 1$ sets an upper bound on $N$:
$N\alpha ^{2}\left( bt\right) ^{4} \ll 1$, which depends on the time
range $t$ we wish to explore. Alternatively speaking,
$(N^B_{flip})^2/N \ll 1$ sets an upper bound on the time range $t$
we can explore for some fixed $N$ using the pseudo-spin model. We
illustrate this bound using the following two examples.

1. If we wish to calculate the Hahn echo signal using the
pseudo-spin model, we shall have $(\alpha N b^2 T_H^2)^2/N \ll 1$
where the Hahn echo decay time $T_H \approx b^{-1/2}
\mathcal{A}^{-1/2} N^{1/4}$ \cite{Espin_YLS}. Therefore, the upper
bound on $N$ is given by $N^2 \alpha^2 b^2 \mathcal{A}^{-2} \ll 1$.
For GaAs quantum dot, this condition is well satisfied for $N
\lesssim 10^8$.

2. For bath of an intermediate size in the allowed region of
$\min[\sqrt{N},N^4 b^2 \Omega^4 \mathcal{A}^{-6}] \gg 1 \gg \alpha^2
N^2b^2\mathcal{A}^{-2}$, e.g., a quantum dot of typical size $N\sim
10^{5} - 10^{6}$ in our problem, $(\alpha N b^2 t^2)^2/N \ll 1$ is
satisfied for a much longer time range $t \sim 10 T_H \sim 100\mu
s$.

In summary, for the pair-correlation approximation (or pseudo-spin
model) to be valid, nuclear spin dynamics of local pair-flips
imposes an upper bound on $N$ while nuclear spin dynamics of
non-local pair-flips imposes a lower bound. Within this mesoscopic
regime, the pair-correlation approximation is well justified. The
error estimation is based on characterizing the difference in the
Hilbert space structure of the exact dynamics and that of the
pseudo-spin model and assuming this difference has a full influence
on the electron spin coherence calculation. Therefore, the bound is
not necessarily tight and it is possible that the pulse control
methodology developed using the pseudo-spin model have actually a
much larger validity regime. Investigation is underway.


\begin{thebibliography}{26}
\bibitem{e-n coupling} The orbital wavefunction of the electron covers all
$N$ nuclei, so the contact hyperfine coupling with each nuclear spin
is $\mathcal{A}/N$, inverse proportional to the bath size $N$.
\end{thebibliography}
\end{document}